%% file: indepEdmond_art_v03.tex
\documentclass[a4paper,10pt]{article}
\usepackage[utf8]{inputenc}
\usepackage[affil-it]{authblk}

\usepackage{graphicx}
\usepackage{wrapfig}
\usepackage{lscape}
\usepackage{rotating}
\usepackage{longtable}
\usepackage{pdflscape}

\usepackage{url}

\usepackage[round]{natbib}
\bibpunct{(}{)}{;}{a}{}{,} 
\title{Independent identification of meteor showers\\ in EDMOND database}
\author{R. Rudawska$^1$, P. Matlovi\v{c}$^1$, J. T\'{o}th$^1$, L. Korno\v{s}$^1$}
\affil{$^1$Faculty of Mathematics, Physics and Informatics, Comenius University, Mlynsk\'{a} dolina, Bratislava, SK-84248, Slovakia}
\date{Dated: \today}

\begin{document}
\maketitle
\begin{abstract}
Cooperation and data sharing among national networks and International Meteor Organization Video Meteor Database (IMO VMDB) resulted in European viDeo MeteOr Network Database (EDMOND). The current version of the database (EDMOND 5.0) contains 144 751 orbits collected from 2001 to 2014. In our survey we used EDMOND database in order to identify existing and new meteor showers in the database.

In the first step of the survey, using $D_{SH}$ criterion we found groups around each meteor within similarity threshold. Mean parameters of the groups were calculated and compared using a new function based on geocentric parameters ($\lambda$, $\alpha$, $\delta$, and $V_g$). Similar groups were merged into final clusters (representing meteor showers), and compared with IAU Meteor Data Center list of meteor showers. This paper presents the results obtained by the proposed methodology.
\parskip=5pt

\end{abstract}

\section{Introduction}
Nowadays, due to the international cooperation, meteor activity is monitored over almost the entire Europe. Consequently, in recent years, multi-national networks of video meteor observers have contributed many new data. As a result, the latest version of EDMOND database contains 144 751 orbits collected from 2001 to 2014.\\

In this paper, we focus on determining an independent method to associate an individual meteor in the EDMOND database with a given meteor shower. The outcome of this method is confirmation of some of the previously reported meteoroid streams listed in the IAU Meteor Data Center (IAU MDC), and finding potentially new ones.\\

In Section \ref{sec:Methodology} we provide necessary mathematical tools used in the independent identification procedure described in Section \ref{sec:IdentProc}. Section \ref{sec:DataPrep} focuses on the EDMOND data preparation used in the analysis. While in Section \ref{sec:Conclusion} we present our conclusions and perspectives for future work.
\section{Methodology}
\label{sec:Methodology}
Our cluster identification procedure links two types of meteor parameters: orbital elements ($e$, $q$, $i$, $\omega$, and $\Omega$) and geocentric parameters ($\lambda$, $\alpha$, $\delta$, and $V_g$). The first set of parameters is applied by so called D-criteria that determine similarity between orbits of meteoroids. While the second set of parameters measure similarity between meteors on the sky in a given meteor shower activity period.

\subsection{Orbital similarity functions}
\label{subsec:Func}
The similarity between two orbits is established by measuring the distance between them with D-criterion (a similarity function). Depending on the number of parameters that defines the similarity function, the distance
between two orbits might be measured in a five- \citep{Southworth_1963, Drummond_1981, Jopek_1993}, seven- \citep{Jopek_2008}, or other dimensional phase.

In our survey, we use two functions. \citet{Southworth_1963} criterion defined as
\begin{eqnarray}
D_{SH}^{2} & = & [e_{B}-e_{A}]^2+[q_{B}-q_{A}]^{2}+\left[2\cdot\sin\frac{I_{AB}}{2}\right]^{2} \nonumber \\ 
 & + & \left[\frac{e_{B}+e_{A}}{2}\right]^{2}\left[2\cdot\sin\frac{\pi_{AB}}{2}\right]^{2},
\label{eq:dsh}
\end{eqnarray}
where $e_A$ and $e_B$ is the eccentricity, and $q_A$ and $q_B$ is the perihelion distance of two orbits, $I_{AB}$ is the angle between two orbital planes, and $\pi_{AB}$ is the distance of the longitudes of perihelia measured
from the intersection of the orbits.

\subsection{Geocentric similarity function}
The second criterion we propose a new distance function $D_x$ involving geocentric parameters, defined as
\begin{eqnarray}
D_x^{2} & = & w_\lambda \left(2\cdot\sin\frac{(\lambda_A-\lambda_B)}{2}\right)^{2}\nonumber \\
        & + & w_\alpha\, (|V_{g_A}-V_{g_B}|+1) \left(2\cdot\sin\left(\frac{\alpha_A-\alpha_B}{2}\cdot\cos\delta_A \right)\right)^{2} \nonumber \\ 
 & + &        w_\delta\, (|V_{g_A}-V_{g_B}|+1) \left(2\cdot\sin\left(\frac{\delta_A-\delta_B}{2}\right)\right)^{2} \nonumber \\ 
 & + &	      w_V\, \left(\frac{|V_{g_A}-V_{g_B}|}{V_{g_A}}\right)^2,
\label{eq:dx}
\end{eqnarray}
where $\lambda_A$ and $\lambda_B$ is the solar longitude, $\alpha_A$ and $\alpha_B$ is the right ascension, $\delta_A$ and $\delta_B$ is the declination, and $V{g_A}$ and $V{g_B}$ is the geocentric velocity of two meteors. The $w_\lambda$, $w_\alpha$, $w_\delta$, and $w_V$ are suitably defined weighting factors. To normalize contribution of each term in $D_x$, we used values: $w_\lambda = \mathrm{0.17}$, $w_\alpha = \mathrm{1.20}$, $w_\delta = \mathrm{1.20}$, and $w_v = \mathrm{0.20}$. Moreover, the values of weighting factors fulfil assumption that compared geocentric parameters differ only 20$^\circ$, 3.5$^\circ$, 3.5$^\circ$, and 3.5 km/s in solar longitude, right ascension, declination and velocity, respectively.

\subsection{Mean parameters}
\label{subsec:MeanCalc}
The mean values of the orbital elements and other parameters of each found cluster were obtained as a weighted arithmetic mean, where the weights were determined by \citet{Welch_2001}
\begin{equation}
 w_i = 1 - \frac{D_{SH}^2}{D_c^2},
\label{eq:weights}
\end{equation}
and where $D_c$ is the threshold of the dynamical similarity.

The mean and standard deviation of angular elements were calculated according to \citet{Mardia_1972}. The mean value of the angular element $\epsilon$ is taken as the solution of the system of equations
\begin{equation}
 \begin{array}{l}
  S = r\, \sin \epsilon\, ,\\ 
  C = r\, \cos \epsilon\, .
 \end{array}
\end{equation}
Here
\begin{equation}
  \begin{array}{lll}
    S = \frac{\sum\limits_{i=1}^N w_i\, \sin\epsilon_i}{\sum\limits_{i=1}^N w_i}\, ,\qquad &
    C = \frac{\sum\limits_{i=1}^N w_i\, \cos\epsilon_i}{\sum\limits_{i=1}^N w_i}\, ,\qquad & 
    r = \sqrt{S^2+C^2}.
  \end{array}
\end{equation}
where $N$ is the number of members in a group/cluster, and the values of the weights $w_i$ are given by Eq. \ref{eq:weights}.

 \begin{figure}[!t]
  \centering
  \includegraphics[trim = 0mm 0mm 0mm 0mm, clip, width=.9\textwidth]{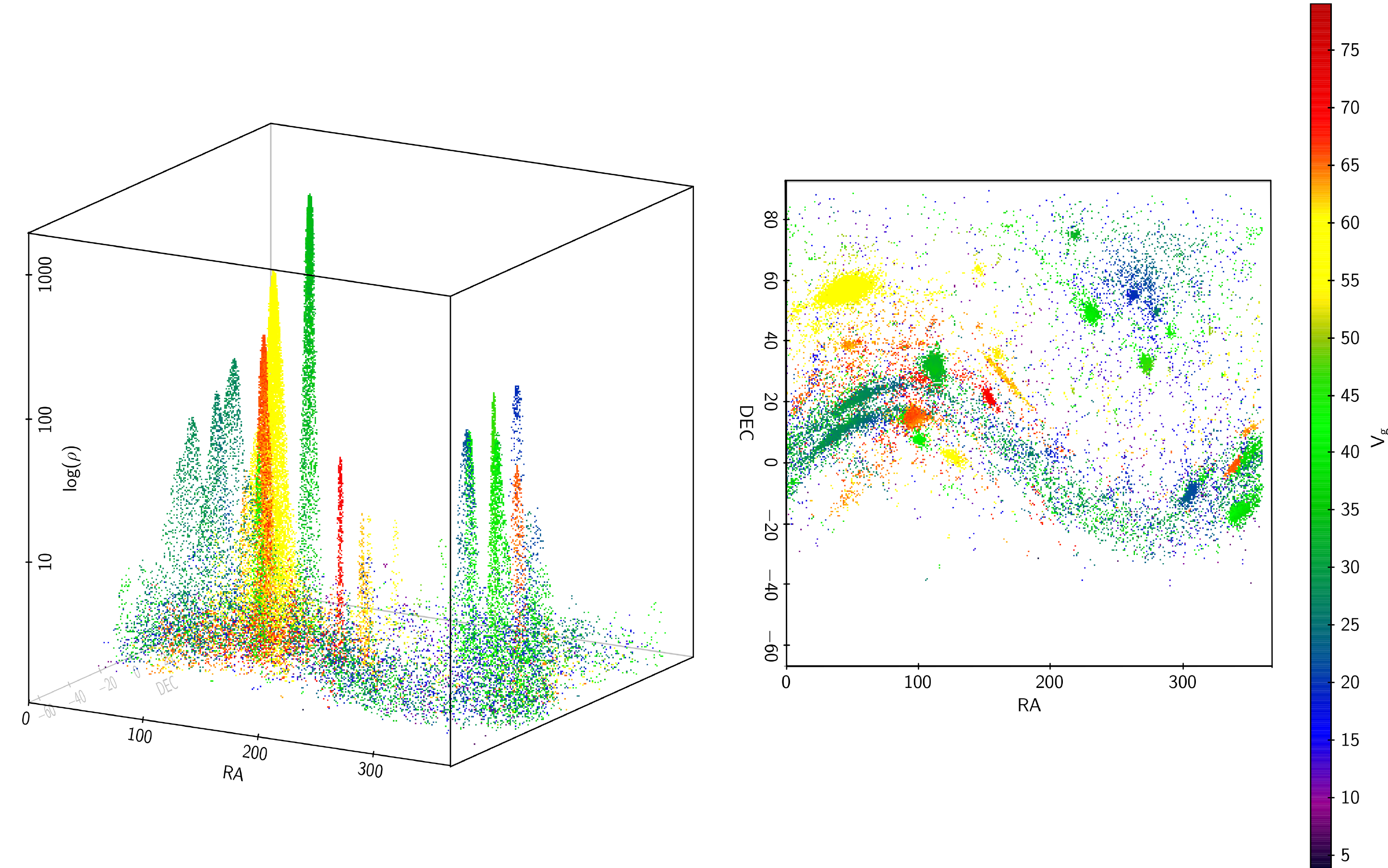}
  \caption{Found groups of orbits within assumed threshold plot in RA, DEC, $V_g$ and $\log\,\rho$, where $\rho$
represents meteor orbits concentrations in the phase space of orbital elements~(Eq. \ref{eq:rho}).}
  \label{fig:rho}
 \end{figure}
\section{Data preparation}
\label{sec:DataPrep}
The current version of the database,~EDMOND 5.0, which contains 144 751 orbits collected from 2001 to 2014, has been split between particular years of observations. At first, we pre-ordered dataset of each year in a way that the starting orbit is with the highest orbits concentrations in the phase space of orbital elements. For this purpose we calculated $\rho$, as defined by Eq. \ref{eq:rho}. As a result, for each year our input data is ordered from the highest to the lowest density $\rho$ (Figure~\ref{fig:rho}).

\section{Identification procedure}
\label{sec:IdentProc}
Our method may be summarised by following steps: 
\begin{description}
 \itemsep=2pt
 \item[Step~1:] We probe database using $D_{SH}$ with a low threshold value $D_c=\mathrm{0.05}$.
    Around a meteoroid orbit is created a sphere of orbital parameters and radius $D_c$. A set of orbits within the sphere creates a group, which members are excluded from following search around another meteoroid orbit. In this way, we have independent groups around each reference meteoroid orbit. Next, for each group a weighted mean of parameters is calculated (Eq.~\ref{eq:weights}).

 \item[Step~2:] Using $D_x$ we are merging groups into clusters of similar weighted means of geocentric parameters found in Step~1. Groups are associated if $D_x \leq D_c'$, where $D_c' = \mathrm{0.15}$.
 
 To calculate mean of parameters of a new cluster, first we search for an orbit within the cluster with the highest density at a point in orbital elements space \citep{Welch_2001}
  \begin{equation}
    \rho= \sum\limits_{i=1}^{N} \left(1 - \frac{D_{i}^2}{D_c^2}\right),
    \label{eq:rho}
  \end{equation}
 where $D_{i}$ is the value of $D_{SH}$ obtained for the $i$-th meteor in the cluster by comparing its orbit with orbits of each member of the cluster, and $D_c$ is the threshold value adjust to a studied cluster. The orbit with the highest $\rho$ is a reference to calculate the new weighted mean of parameters for cluster found in Step~2.
  
 We repeat Step 2 using new means till groups are no longer linked into clusters.
 \item[Step 3:] We compare parameters of known meteor showers in the~IAU MDC with the final mean values of the same parameters of found clusters. For this purpose we use $D_{SH}$ criterion with $D_c = \mathrm{0.15}$.
 We merge clusters of the same identified meteor shower. Although, a cluster must have 5 or more members to be considered as a representation of a meteor shower. 
\end{description}
\section{Results}
\label{sec:Results}
\begin{figure}[h]
 \centering
 \includegraphics[trim= 0mm 0mm 0mm 0mm, clip, width=.32\textwidth]{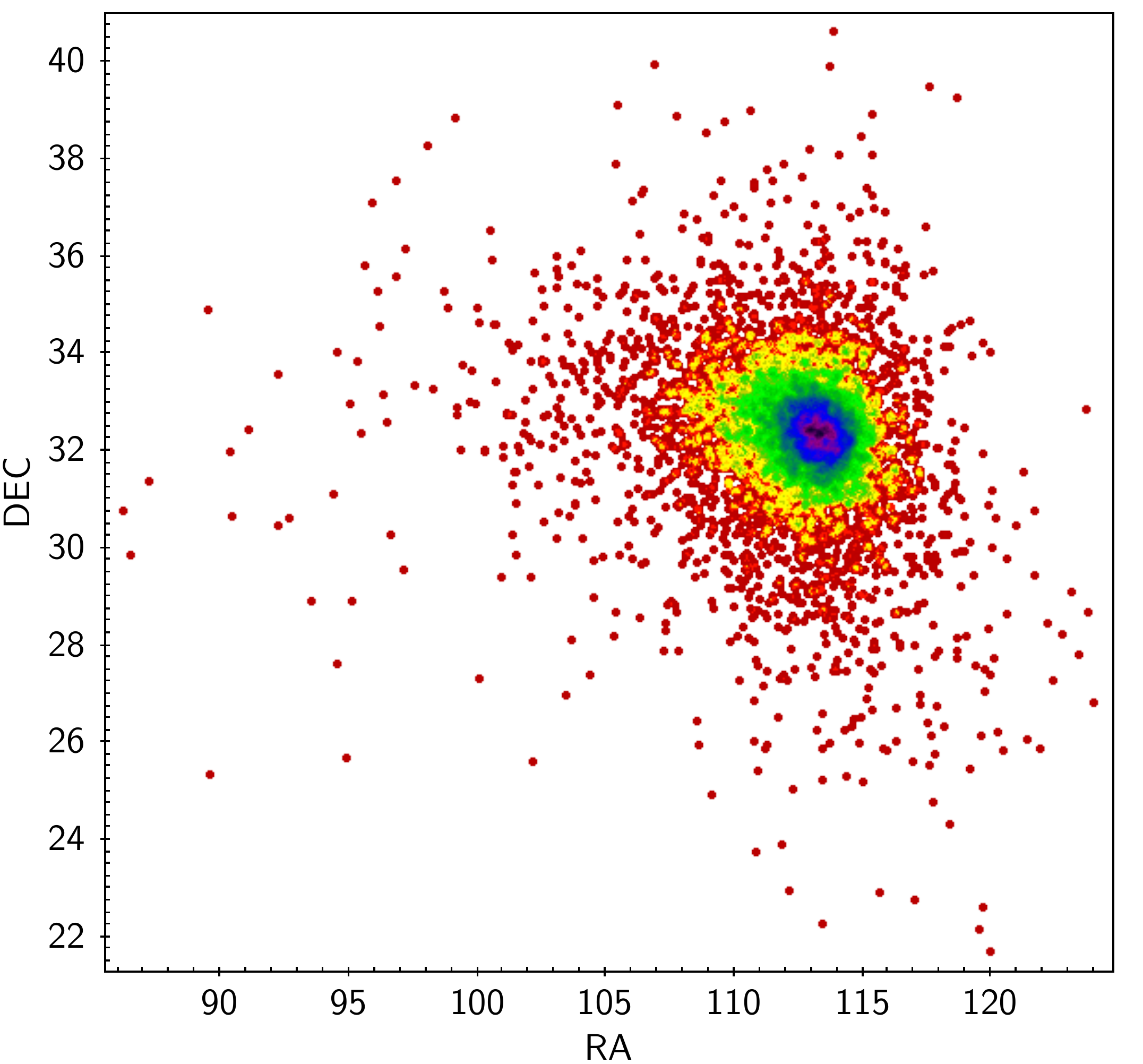}
 \includegraphics[trim= 0mm 0mm 0mm 0mm, clip, width=.32\textwidth]{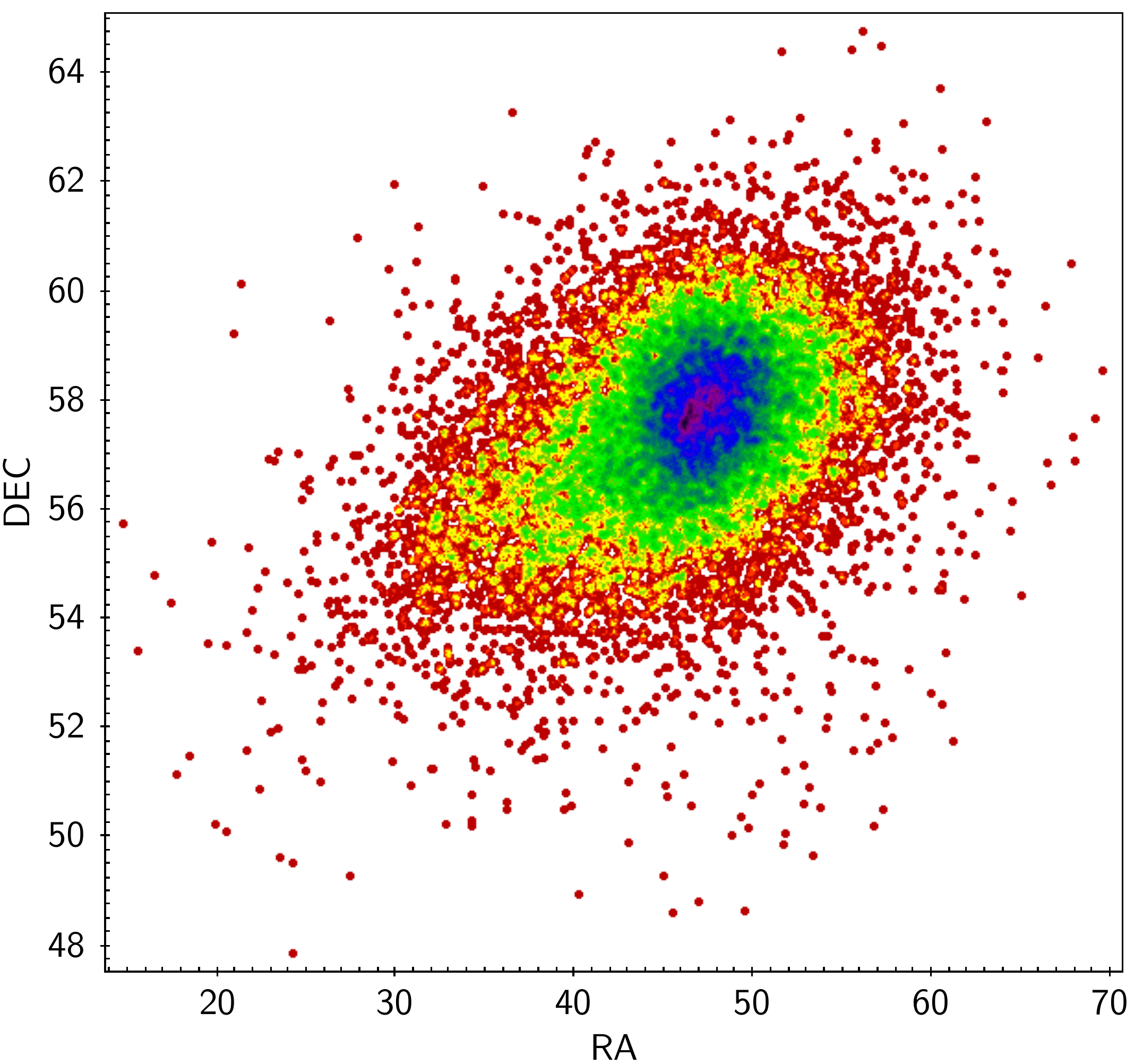}
 \includegraphics[trim= 0mm 0mm 0mm 0mm, clip, width=.32\textwidth]{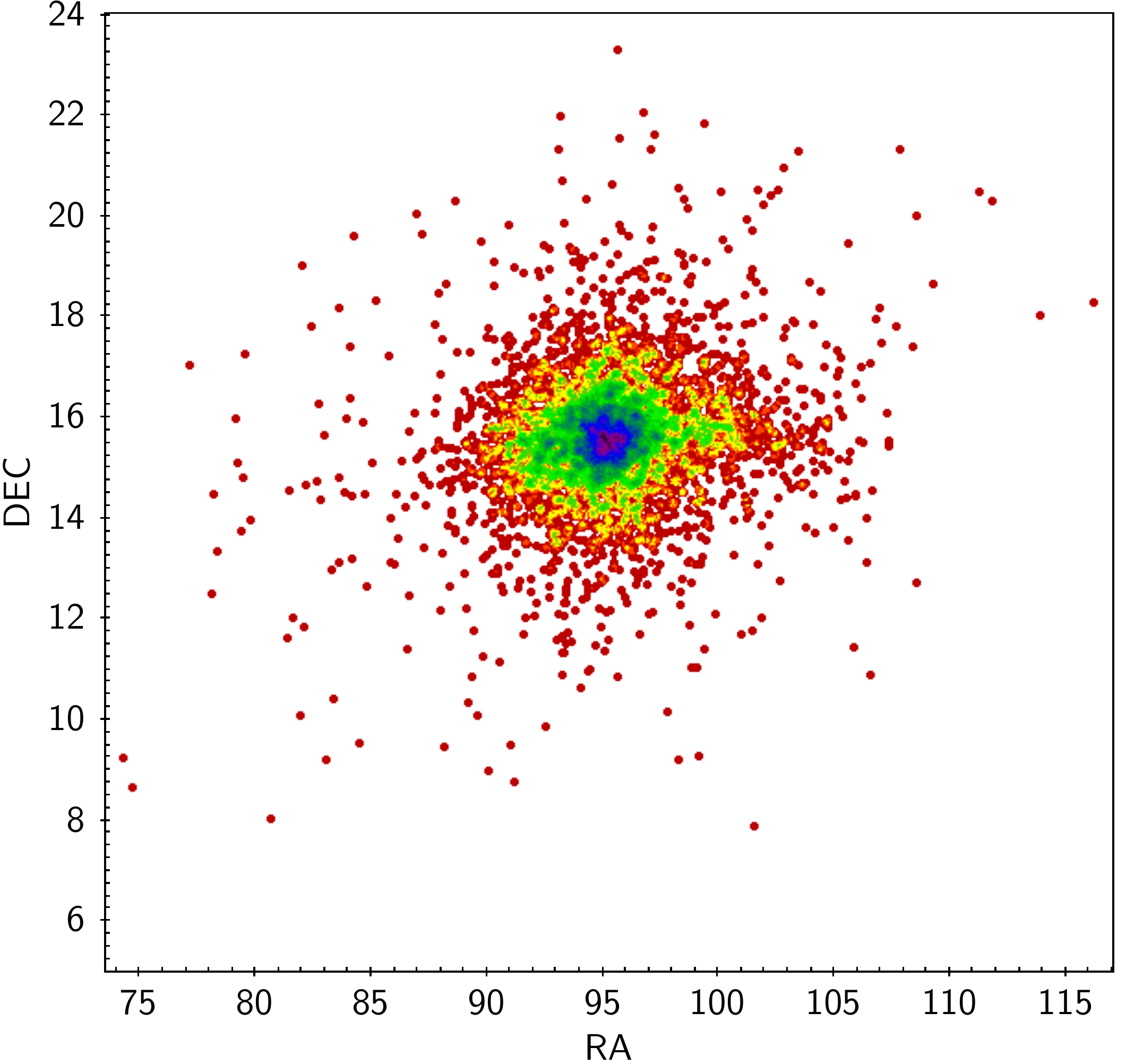}
 \caption{Identified meteor showers: Geminids (left), Perseids (centre), and Orionids (right). Colours represent meteor concentrations on the sky.}
 \label{fig:GEM_PER_ORI}
\end{figure}

\begin{figure}[h]
 \centering
 \includegraphics[trim= 0mm 0mm 0mm 0mm, clip, width=.32\textwidth]{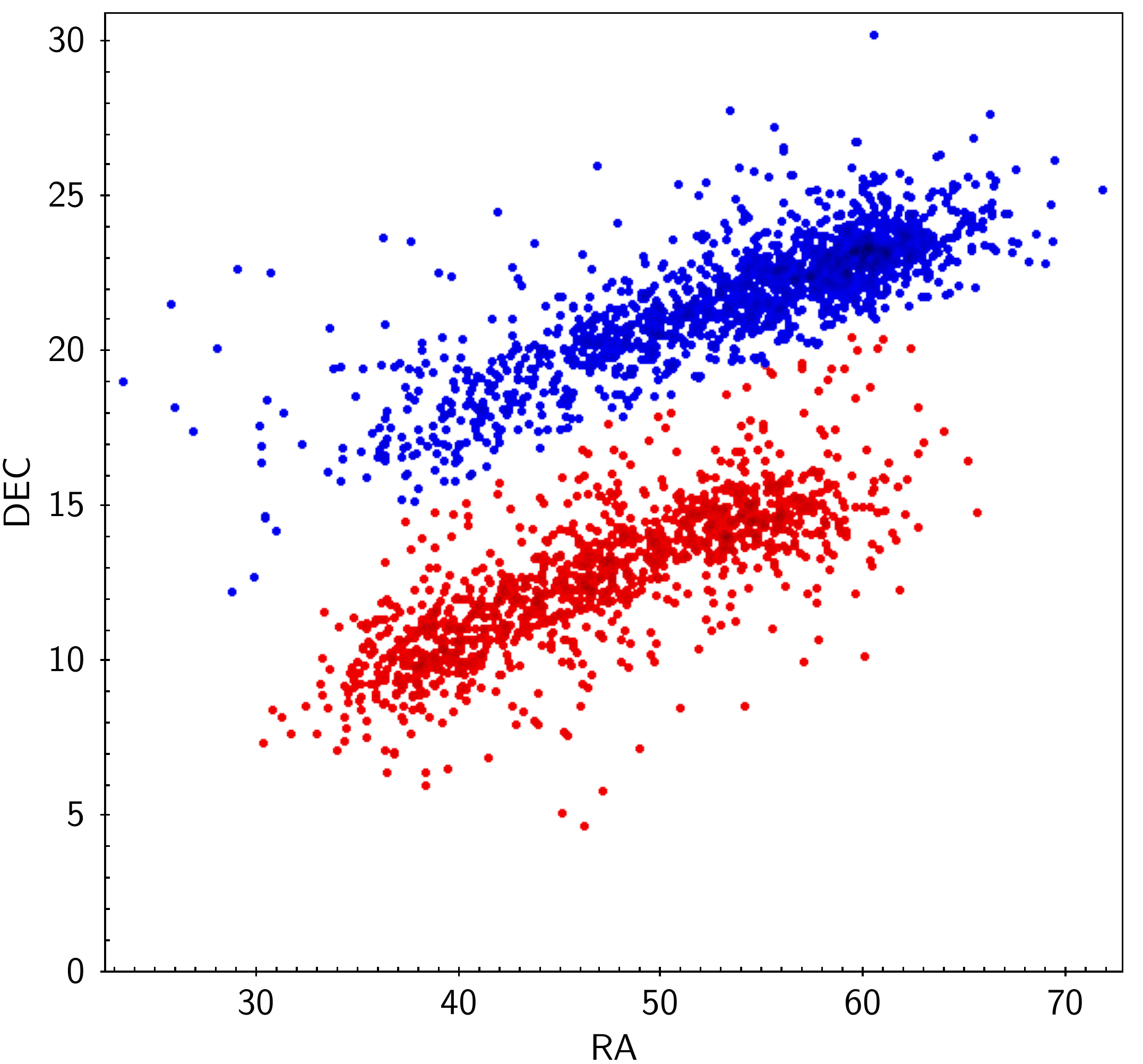}
 \includegraphics[trim= 0mm 0mm 0mm 0mm, clip, width=.32\textwidth]{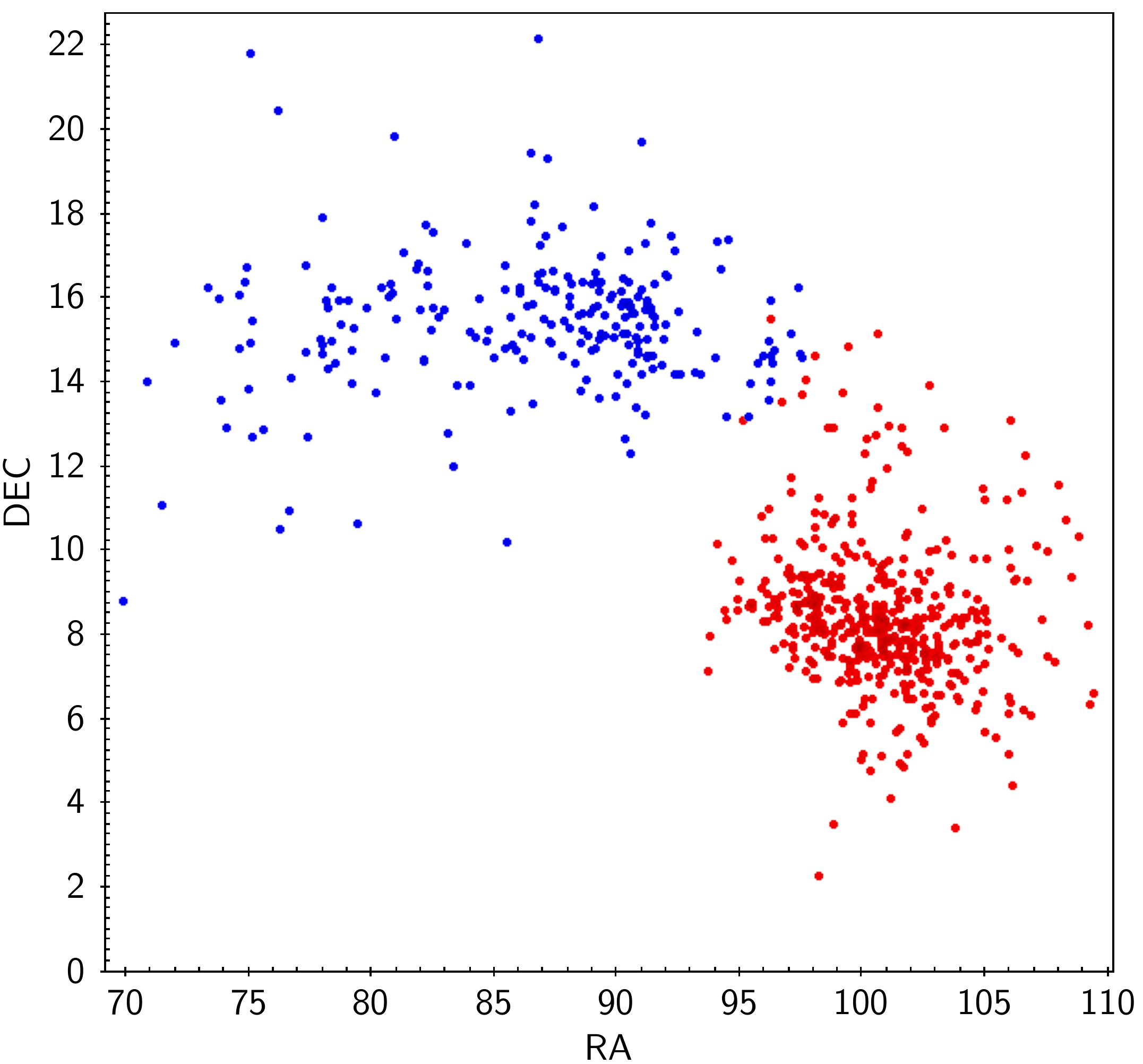}
 \includegraphics[trim= 0mm 0mm 0mm 0mm, clip, width=.32\textwidth]{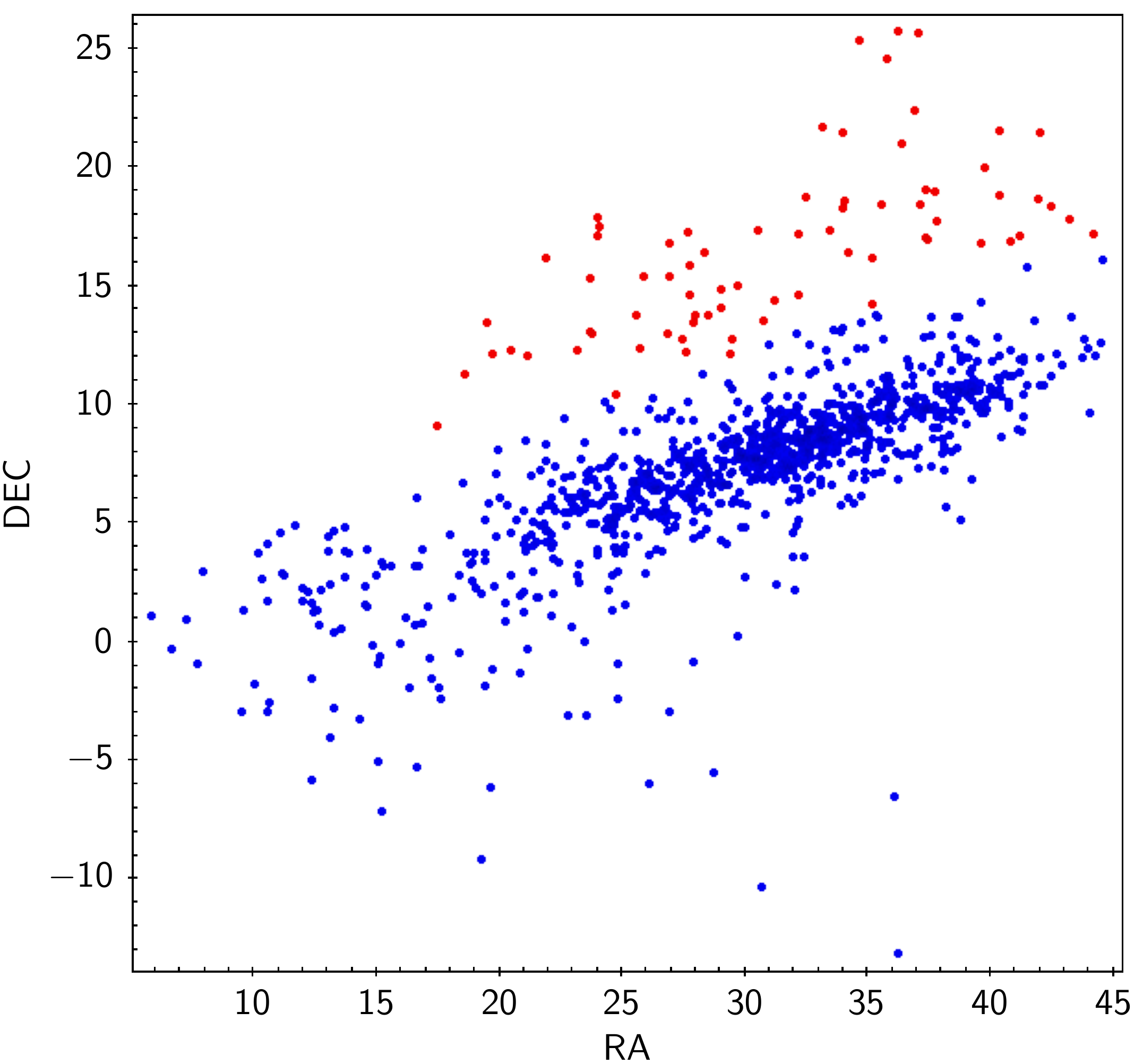}
 \caption{Identified meteor showers: Southern \& Northern Taurids (left), December Monocerotids \& November Orionids (centre), and Northern \& Southern October $\delta$ Arietids (right). Colours indicated different showers of the pair -- respectively red \& blue.}
 \label{fig:pairs}
\end{figure}

\begin{sidewaysfigure}[]
 \centering
 \includegraphics[trim= 0mm 0mm 0mm 0mm, clip, width=1.\textwidth]{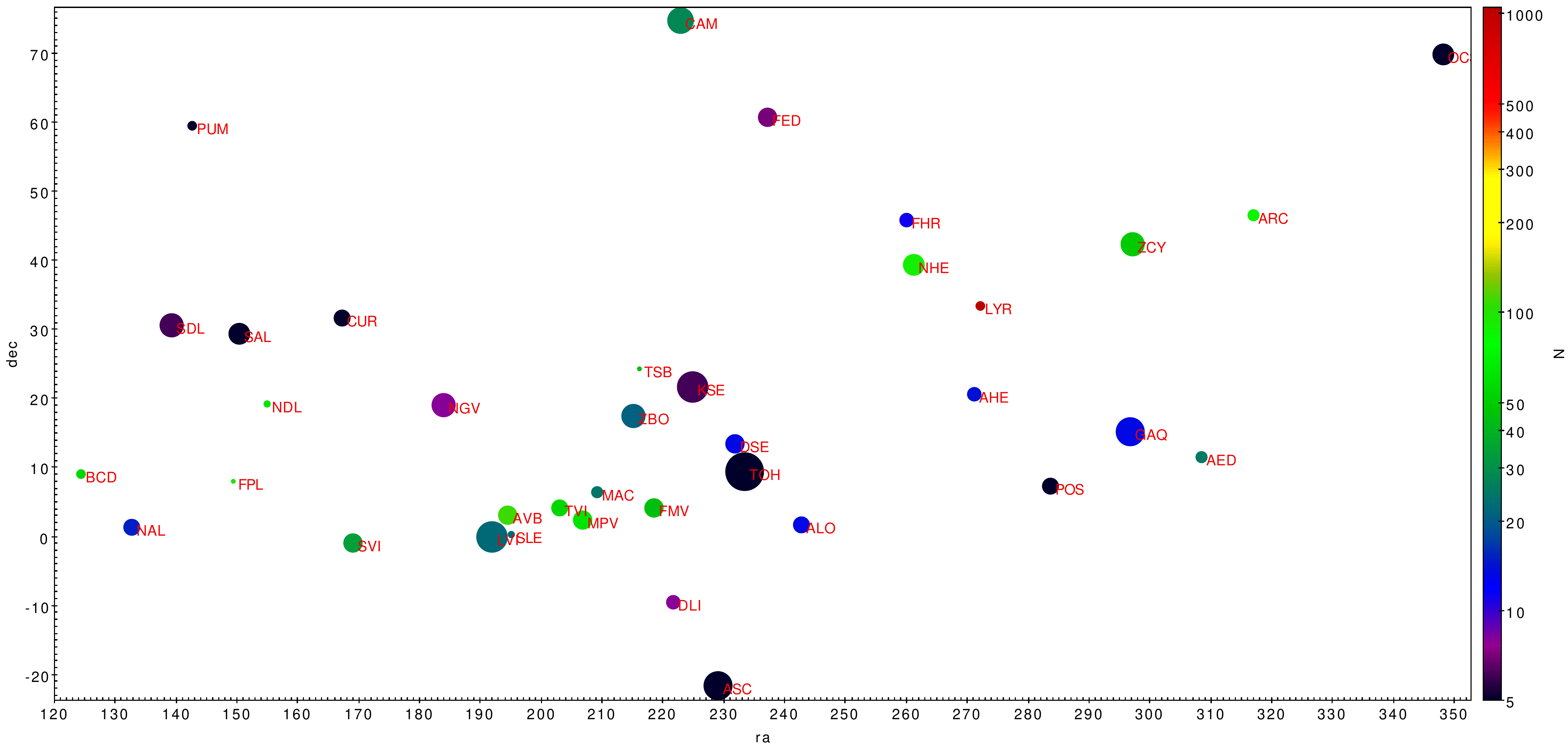}
 \caption{Identified spring meteor showers. Colour represents number of members of a shower, while the size represents $D$ value based on the similarity measure between the mean orbital parameters of a cluster and linked with it the IAU MDC meteor shower (column 12 and 13 in Table~\ref{tab:meanOrbRad}, respectively).}
 \label{fig:spring}
\end{sidewaysfigure}

\begin{sidewaysfigure}[]
 \centering
 \includegraphics[trim= 0mm 0mm 0mm 0mm, clip, width=1.\textwidth]{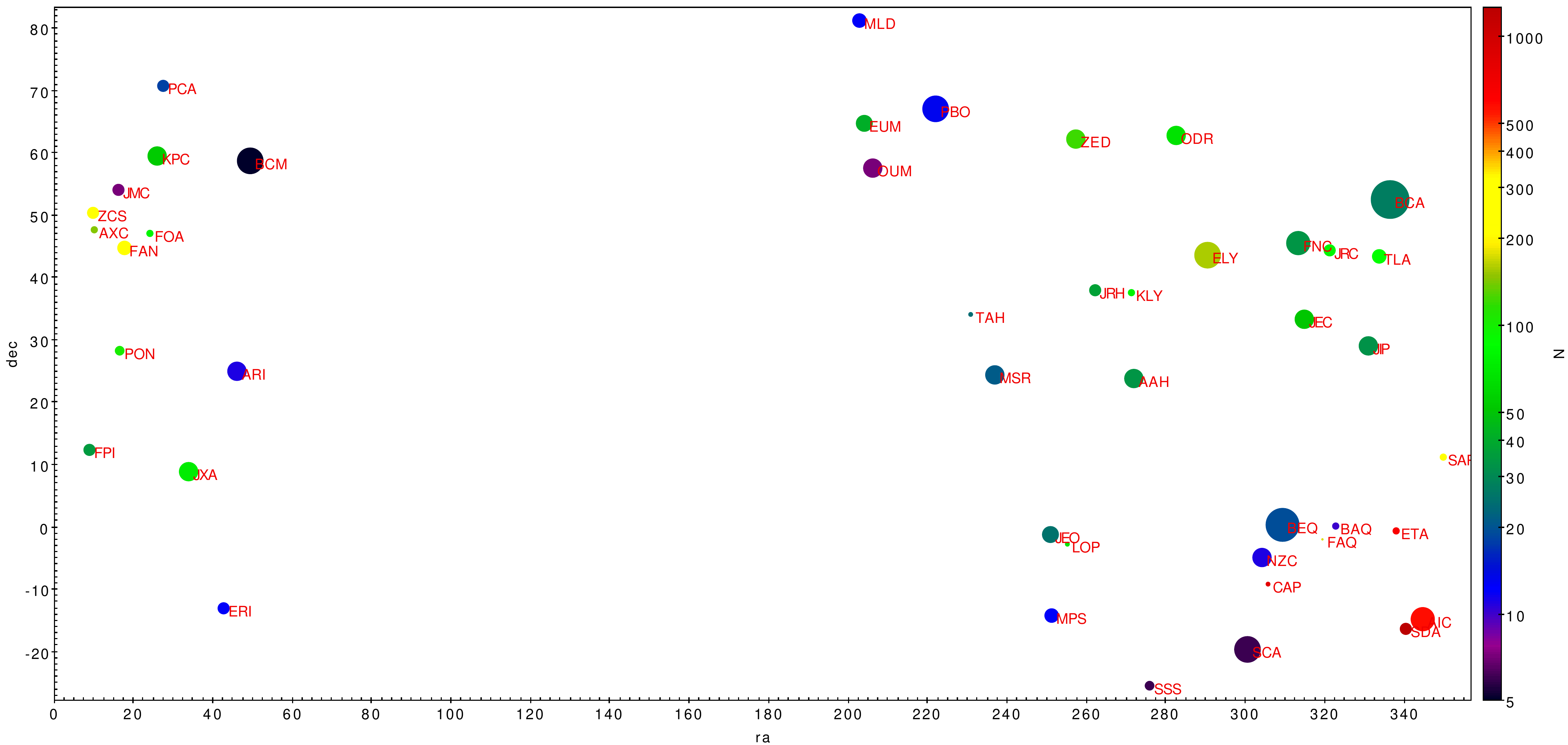}
 \caption{Identified summer meteor showers. Colour represents number of members of a shower, while the size represents $D$ value based on the similarity measure between the mean orbital parameters of a cluster and linked with it the IAU MDC meteor shower (column 12 and 13 in Table~\ref{tab:meanOrbRad}, respectively).}
 \label{fig:summer}
\end{sidewaysfigure}

\begin{sidewaysfigure}[]
 \centering
 \includegraphics[trim= 0mm 0mm 0mm 0mm, clip, width=1.\textwidth]{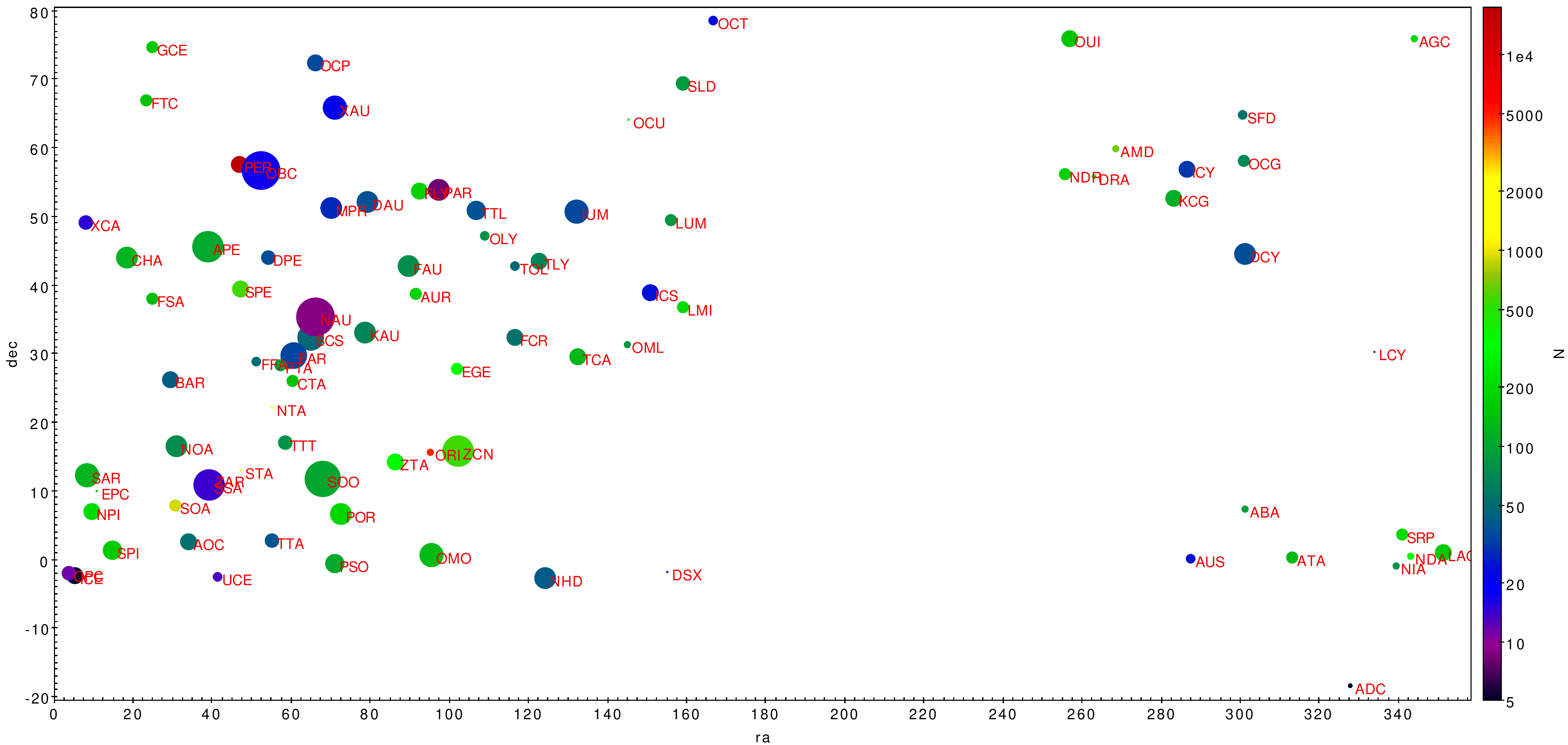}
 \caption{Identified autumn meteor showers. Colour represents number of members of a shower, while the size represents $D$ value based on the similarity measure between the mean orbital parameters of a cluster and linked with it the IAU MDC meteor shower (column 12 and 13 in Table~\ref{tab:meanOrbRad}, respectively).}
 \label{fig:autumn}
\end{sidewaysfigure}

\begin{sidewaysfigure}[]
 \centering
 \includegraphics[trim= 0mm 0mm 0mm 0mm, clip, width=1.\textwidth]{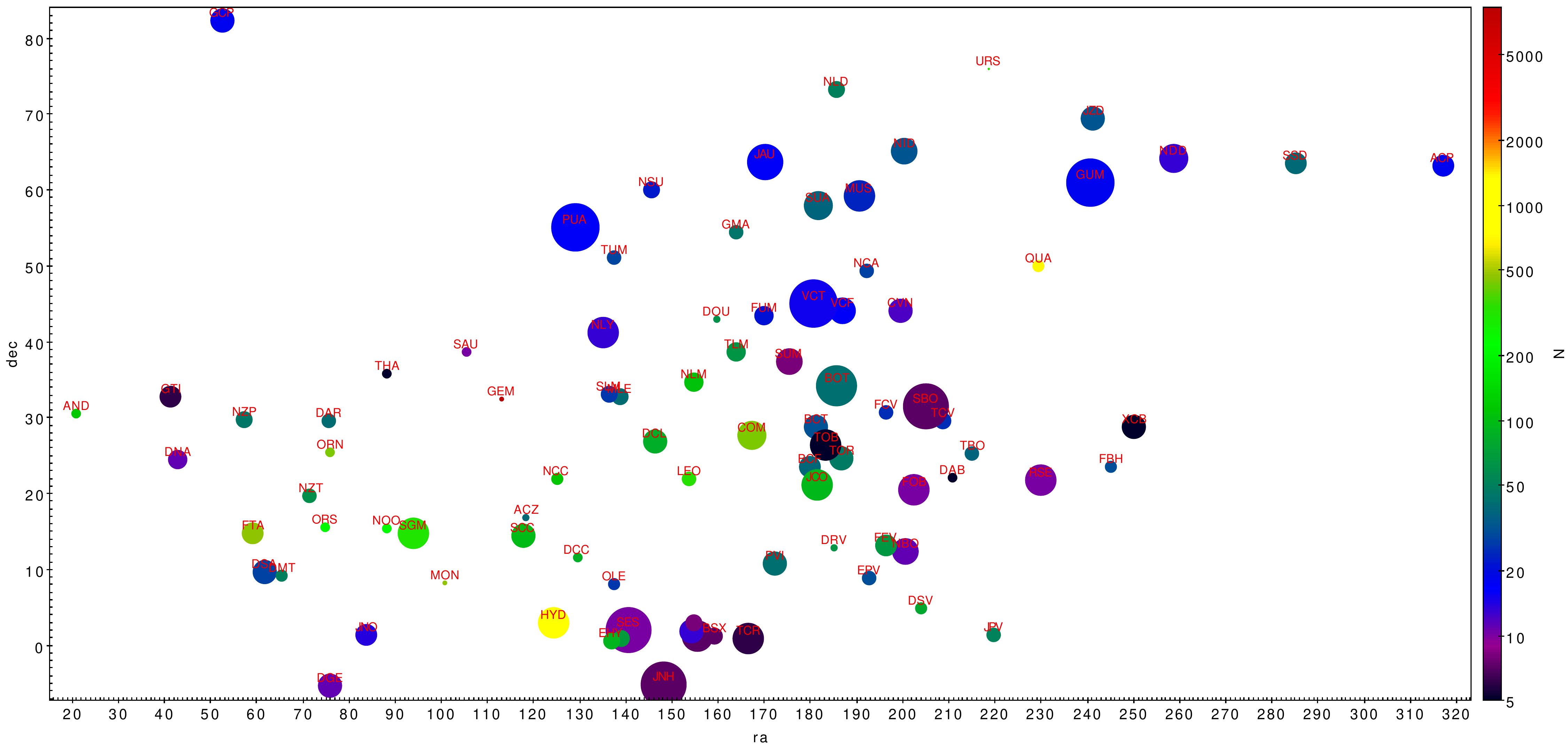}
 \caption{Identified winter meteor showers. Colour represents number of members of a shower, while the size represents $D$ value based on the similarity measure between the mean orbital parameters of a cluster and linked with it the IAU MDC meteor shower (column 12 and 13 in Table~\ref{tab:meanOrbRad}, respectively).}
 \label{fig:winter}
\end{sidewaysfigure}

The results of our survey are given in Table \ref{tab:meanOrbRad} and Figure~\ref{fig:GEM_PER_ORI}-\ref{fig:winter}. It contains 257 meteoroid streams identified by described earlier procedure. It summarizes the mean geocentric parameters and mean orbital parameters of the detected showers (clusters), ranked according to the IAU MDC coding. In addition, in the last column, is given value of $D_{SH}$ that determines the similarity between the mean orbit of a cluster and the orbit of a given meteoroid stream from the~IAU~MDC. Additionally, we visualise results separately for each seasons in Figures~\ref{fig:spring}-\ref{fig:winter}, where colours represent amount of meteors within an identified meteor shower, while the size of points corresponds to $D_{SH}$ value.\\

To show efficiency of the procedure we present here results for selected cases. Figure~\ref{fig:GEM_PER_ORI} shows meteor concentrations of Geminids, Perseids, and Orionids on the sky. Those meteor showers are the most prominent showers in the EDMOND database, including over 5 000 members. Those showers have been correctly identified by our procedure. Their activity period lasts about 25-35 days. As should be expected, Geminids and Orionids are more compact in comparison to Perseids meteor shower. But in contrast, Perseids is more prominent than the other two showers.\\

A given identification method may fail in separation of branches of the same meteor shower. Moreover, if two meteor showers are located in close distance to each other on the sky, an identification method based on geocentric parameters may fail and link those two showers into one. However, our identification procedure succeeds in correct separation of meteor showers in such cases. Figure~\ref{fig:pairs} presents example of such pairs as: Southern \& Northern Taurids, December Monocerotids \& November Orionids, and Northern \& Southern October $\delta$ Arietids. The second listed shower of a given pair is marked in blue. As shown in Figure~\ref{fig:pairs} our identification procedure correctly separates meteor showers.

\section{Conclusion}
\label{sec:Conclusion}
In order to apply $D_{SH}$ criterion in Step~3 for identification of clusters, we selected meteor showers for which their orbital elements are provided by the IAU MDC (as of June 2014). In total we used 488 meteor showers.\\

We identified 257 meteor showers. The list includes 42 already established streams, 152 from the working list and 63 \textit{pro-tempore} meteor showers. For a~higher threshold ($D_{c}'' = \mathrm{0.20}$), we found 284 meteor shower in total (44, 173, and 67, respectively). However, with such threshold value some of the showers are more contaminated by the sporadic background.\\

There are several clusters that require further investigation. Some of them are those meteor showers for which orbital elements are not given at the IAU MDC. We plan to identify them using $D_x$ criterion, calculate their orbital elements which will be provided to the IAU MDC subsequently. Of course, not identified yet clusters may represents also possible new meteor showers, which need our additional, detailed analysis before they will be submitted to the IAU MDC as well.
\section{Acknowledgement}
The work is supported by the Slovak grant APVV-0517-12, APVV-0516-10 and VEGA 1/0225/14.
\nocite{*}
\bibliographystyle{aa}
\bibliography{indepEdmond}
\begin{center}
 \small
 \input{estwork_SH0.15}
 \normalsize
\end{center}

\end{document}

%% file: estwork_SH0.15.tex
\begin{longtable}{rr|rrrr|rrrrr|r|r}
 \label{tab:meanOrbRad}\\
 \hline
  IAU & Code &  $\lambda_\odot$ & $\alpha$ & $\delta$ & $V_g$ & $e$ & $q$ & $i$ & $\omega$ & $\Omega$ & No & $D_{SH}$ \\
 \hline
 \endfirsthead
 \multicolumn{13}{r}%
  {\tablename\ \thetable\ -- \textit{Continued from previous page}} \\
 \hline
  IAU & Code &  $\lambda_\odot$ & $\alpha$ & $\delta$ & $V_g$ & $e$ & $q$ & $i$ & $\omega$ & $\Omega$ & No & $D_{SH}$ \\
 \hline
 \endhead
 \hline \multicolumn{13}{r}{\textit{Continued on next page}} \\
 \endfoot
 \hline
 \endlastfoot
001 &  CAP &     127.2 &  305.8 &  -9.3 &  21.84 &      0.748 &      0.601 &        7.1 &      267.5 &      127.2&         807 &  0.03 \\ 
002 &  STA &     215.3 &   47.1 &  12.9 &  27.62 &      0.816 &      0.347 &        5.3 &      116.7 &       35.3&        1155 &  0.02 \\ 
004 &  GEM &     261.7 &  113.1 &  32.4 &  33.51 &      0.885 &      0.147 &       22.5 &      324.3 &      261.7&        8268 &  0.03 \\ 
005 &  SDA &     127.5 &  340.3 & -16.4 &  40.14 &      0.967 &      0.080 &       26.8 &      151.0 &      307.5&        1252 &  0.06 \\ 
006 &  LYR &      32.2 &  272.1 &  33.3 &  46.17 &      0.922 &      0.918 &       79.0 &      215.0 &       32.2&        1045 &  0.06 \\ 
007 &  PER &     139.5 &   46.8 &  57.6 &  58.27 &      0.886 &      0.947 &      112.5 &      149.2 &      139.5&       17265 &  0.10 \\ 
008 &  ORI &     208.1 &   95.2 &  15.6 &  65.49 &      0.907 &      0.568 &      163.5 &       84.7 &       28.1&        4835 &  0.04 \\ 
009 &  DRA &     195.0 &  262.9 &  55.8 &  20.02 &      0.650 &      0.996 &       30.8 &      173.3 &      195.0&         496 &  0.03 \\ 
010 &  QUA &     283.1 &  229.4 &  50.0 &  40.03 &      0.624 &      0.980 &       70.4 &      173.0 &      283.1&        1381 &  0.07 \\ 
012 &  KCG &     136.7 &  283.1 &  52.6 &  23.33 &      0.704 &      0.982 &       36.5 &      202.4 &      136.7&         112 &  0.11 \\ 
013 &  LEO &     235.7 &  153.6 &  21.9 &  69.91 &      0.851 &      0.984 &      162.1 &      171.7 &      235.7&         332 &  0.08 \\ 
015 &  URS &     270.4 &  218.6 &  75.9 &  32.63 &      0.798 &      0.937 &       52.1 &      206.8 &      270.4&         316 &  0.01 \\ 
016 &  HYD &     255.0 &  124.4 &   2.9 &  58.28 &      0.974 &      0.248 &      128.5 &      121.3 &       75.0&         976 &  0.17 \\ 
017 &  NTA &     225.8 &   55.3 &  22.1 &  27.90 &      0.825 &      0.354 &        2.9 &      294.9 &      225.8&        1403 &  0.01 \\ 
018 &  AND &     226.4 &   20.9 &  30.5 &  16.71 &      0.692 &      0.781 &        9.3 &      241.0 &      226.4&         115 &  0.05 \\ 
019 &  MON &     258.8 &  100.7 &   8.2 &  40.98 &      0.979 &      0.188 &       34.9 &      129.4 &       78.8&         467 &  0.03 \\ 
020 &  COM &     273.8 &  167.2 &  27.7 &  62.61 &      0.922 &      0.540 &      134.9 &      266.5 &      273.8&         435 &  0.16 \\ 
021 &  AVB &      19.0 &  194.5 &   3.0 &  19.32 &      0.709 &      0.693 &        4.5 &      254.8 &       19.0&         109 &  0.13 \\ 
022 &  LMI &     208.2 &  159.2 &  36.7 &  60.76 &      0.930 &      0.599 &      124.7 &      100.1 &      208.2&         201 &  0.07 \\ 
023 &  EGE &     205.1 &  101.8 &  27.8 &  67.80 &      0.885 &      0.765 &      170.7 &      239.9 &      205.1&         330 &  0.07 \\ 
025 &  NOA &     195.7 &   31.0 &  16.5 &  33.31 &      0.909 &      0.193 &        5.8 &      314.1 &      195.7&          74 &  0.13 \\ 
026 &  NDA &     137.9 &  343.0 &   0.5 &  38.06 &      0.954 &      0.102 &       21.4 &      328.4 &      137.9&         344 &  0.05 \\ 
027 &  KSE &       6.3 &  224.9 &  21.7 &  41.28 &      0.951 &      0.494 &       59.4 &      272.4 &        6.3&           6 &  0.21 \\ 
028 &  SOA &     195.0 &   30.8 &   7.8 &  28.06 &      0.821 &      0.307 &        5.5 &      122.8 &       15.6&         903 &  0.08 \\ 
031 &  ETA &      46.0 &  338.1 &  -0.7 &  65.35 &      0.933 &      0.576 &      163.4 &       96.1 &       46.0&         645 &  0.04 \\ 
033 &  NIA &     148.2 &  339.5 &  -0.9 &  28.04 &      0.828 &      0.366 &        8.2 &      294.8 &      148.2&          79 &  0.04 \\ 
034 &  DSE &     318.8 &  231.9 &  13.5 &  63.18 &      0.839 &      0.962 &      126.3 &      193.9 &      318.8&          13 &  0.13 \\ 
038 &  CUR &     350.6 &  167.3 &  31.7 &  17.60 &      0.642 &      0.762 &       12.1 &      243.9 &      350.6&           5 &  0.11 \\ 
039 &  NAL &     342.4 &  132.7 &   1.4 &  11.43 &      0.576 &      0.919 &        4.6 &       38.1 &      162.4&          15 &  0.11 \\ 
040 &  ZCY &      12.7 &  297.1 &  42.3 &  41.26 &      0.807 &      0.916 &       70.4 &      144.5 &       12.7&          49 &  0.15 \\ 
047 &  DLI &      30.5 &  221.8 &  -9.5 &  29.75 &      0.849 &      0.349 &        7.7 &      294.9 &       30.5&           8 &  0.10 \\ 
049 &  LVI &       2.5 &  191.9 &  -0.1 &  26.65 &      0.796 &      0.432 &        5.2 &      284.9 &        2.5&          22 &  0.20 \\ 
052 &  OUM &      68.0 &  206.3 &  57.5 &  12.07 &      0.550 &      1.013 &       16.6 &      182.6 &       68.0&           7 &  0.10 \\ 
055 &  ASC &      37.8 &  229.0 & -21.7 &  32.36 &      0.892 &      0.285 &        5.2 &      122.0 &      217.8&           5 &  0.19 \\ 
061 &  TAH &      75.1 &  230.9 &  34.1 &  14.71 &      0.622 &      0.965 &       19.3 &      208.9 &       75.1&          24 &  0.03 \\ 
083 &  OCG &     193.4 &  300.9 &  58.1 &  19.71 &      0.646 &      0.977 &       29.4 &      198.6 &      193.4&          70 &  0.08 \\ 
088 &  ODR &     109.8 &  282.7 &  62.7 &  27.84 &      0.711 &      1.011 &       45.6 &      188.7 &      109.8&          66 &  0.10 \\ 
089 &  PVI &     286.0 &  172.2 &  10.8 &  63.52 &      0.887 &      0.423 &      161.4 &      282.4 &      286.0&          42 &  0.13 \\ 
090 &  JCO &     290.1 &  181.4 &  21.2 &  61.60 &      0.892 &      0.500 &      135.2 &      272.4 &      290.1&          94 &  0.18 \\ 
096 &  NCC &     292.9 &  125.2 &  22.0 &  27.32 &      0.815 &      0.405 &        2.7 &      287.5 &      292.9&         113 &  0.07 \\ 
097 &  SCC &     291.8 &  117.9 &  14.5 &  25.51 &      0.791 &      0.469 &        5.2 &      100.3 &      111.8&          96 &  0.14 \\ 
112 &  NDL &     334.3 &  155.0 &  19.2 &  21.50 &      0.739 &      0.638 &        5.6 &      260.6 &      334.3&          62 &  0.04 \\ 
113 &  SDL &     327.5 &  139.2 &  30.5 &  16.78 &      0.708 &      0.796 &        7.1 &      238.3 &      327.5&           6 &  0.15 \\ 
124 &  SVI &     350.6 &  169.1 &  -1.0 &  22.58 &      0.753 &      0.575 &        2.8 &       88.5 &      170.6&          34 &  0.12 \\ 
125 &  SAL &     349.6 &  150.5 &  29.3 &  13.77 &      0.649 &      0.873 &        5.1 &      226.1 &      349.6&           5 &  0.14 \\ 
133 &  PUM &      24.8 &  142.7 &  59.4 &   9.56 &      0.563 &      1.003 &        9.9 &      181.8 &       24.8&           5 &  0.06 \\ 
134 &  NGV &      12.2 &  183.9 &  19.0 &  15.67 &      0.665 &      0.826 &        8.4 &      237.5 &       12.2&           8 &  0.16 \\ 
136 &  SLE &      12.5 &  195.1 &   0.3 &  22.78 &      0.747 &      0.552 &        4.9 &      273.5 &       12.5&          22 &  0.04 \\ 
145 &  ELY &      49.5 &  290.4 &  43.5 &  43.51 &      0.910 &      1.001 &       74.1 &      191.1 &       49.5&         157 &  0.14 \\ 
164 &  NZC &      93.9 &  304.3 &  -5.0 &  37.05 &      0.921 &      0.123 &       40.7 &      328.4 &       93.9&          11 &  0.11 \\ 
168 &  SSS &      82.5 &  276.0 & -25.6 &  29.74 &      0.857 &      0.332 &        2.8 &      117.2 &      262.5&           6 &  0.05 \\ 
171 &  ARI &      82.0 &   46.2 &  25.0 &  40.16 &      0.967 &      0.062 &       30.4 &       25.3 &       82.0&          11 &  0.11 \\ 
177 &  BCA &     115.6 &  336.5 &  52.5 &  47.62 &      0.880 &      0.986 &       85.4 &      199.9 &      115.6&          27 &  0.21 \\ 
179 &  SCA &     101.1 &  300.5 & -19.7 &  32.78 &      0.898 &      0.234 &        1.6 &      309.3 &      101.1&           6 &  0.15 \\ 
186 &  EUM &     100.6 &  204.0 &  64.8 &  15.80 &      0.607 &      1.002 &       21.9 &      164.0 &      100.6&          41 &  0.09 \\ 
187 &  PCA &     115.3 &   27.5 &  70.7 &  40.53 &      0.717 &      0.842 &       71.9 &      125.0 &      115.3&          18 &  0.06 \\ 
191 &  ERI &     135.3 &   42.8 & -13.1 &  63.40 &      0.893 &      0.948 &      130.7 &       30.5 &      315.3&          12 &  0.07 \\ 
193 &  ZAR &     138.4 &   39.6 &  11.7 &  69.60 &      0.915 &      0.949 &      174.0 &       29.8 &      318.4&          18 &  0.09 \\ 
194 &  UCE &     148.0 &   41.3 &  -2.6 &  62.23 &      0.801 &      0.650 &      142.8 &       78.4 &      328.0&          13 &  0.06 \\ 
199 &  ADC &     146.5 &  327.9 & -18.5 &  20.93 &      0.739 &      0.619 &        2.8 &       85.1 &      326.5&           5 &  0.03 \\ 
205 &  XAU &     152.1 &   71.0 &  65.8 &  54.61 &      0.872 &      0.934 &      103.1 &      146.9 &      152.1&          21 &  0.15 \\ 
206 &  AUR &     158.6 &   91.4 &  38.7 &  64.75 &      0.925 &      0.652 &      148.6 &      105.5 &      158.6&         184 &  0.07 \\ 
207 &  SCS &     158.7 &   65.1 &  32.4 &  69.31 &      0.881 &      1.003 &      162.4 &      179.3 &      158.7&          47 &  0.16 \\ 
208 &  SPE &     167.2 &   47.1 &  39.4 &  63.43 &      0.910 &      0.705 &      138.5 &      248.2 &      167.2&         568 &  0.10 \\ 
215 &  NPI &     171.8 &    9.7 &   7.0 &  29.41 &      0.853 &      0.285 &        4.7 &      304.4 &      171.8&         206 &  0.10 \\ 
216 &  SPI &     176.8 &   14.7 &   1.4 &  30.07 &      0.861 &      0.266 &        5.3 &      126.9 &      351.3&         160 &  0.12 \\ 
217 &  OPC &     186.4 &    3.9 &  -2.0 &  20.57 &      0.752 &      0.610 &        2.3 &       84.8 &        6.5&          11 &  0.09 \\ 
219 &  SAR &     167.6 &    8.5 &  12.3 &  37.87 &      0.955 &      0.124 &       19.1 &      324.0 &      167.6&         120 &  0.15 \\ 
220 &  NDR &     158.5 &  255.8 &  56.1 &  19.22 &      0.627 &      1.000 &       29.5 &      175.7 &      158.5&         174 &  0.07 \\ 
221 &  DSX &     188.0 &  155.1 &  -1.9 &  31.28 &      0.857 &      0.150 &       23.3 &      212.5 &        8.0&          14 &  0.02 \\ 
224 &  DAU &     185.4 &   79.2 &  52.1 &  62.36 &      0.889 &      0.888 &      127.4 &      221.2 &      185.4&          37 &  0.13 \\ 
226 &  ZTA &     194.8 &   86.5 &  14.2 &  66.30 &      0.871 &      0.704 &      161.8 &       68.5 &       14.8&         305 &  0.10 \\ 
227 &  OMO &     196.6 &   95.4 &   0.6 &  64.39 &      0.818 &      0.916 &      139.1 &       35.5 &       17.7&         132 &  0.15 \\ 
228 &  OLY &     205.7 &  109.0 &  47.1 &  64.13 &      0.864 &      0.893 &      134.9 &      218.6 &      205.7&          79 &  0.06 \\ 
229 &  NAU &     198.6 &   66.3 &  35.3 &  56.58 &      0.966 &      0.196 &      133.1 &      308.8 &      198.6&           9 &  0.24 \\ 
230 &  ICS &     204.2 &  150.7 &  38.9 &  61.46 &      0.923 &      0.643 &      126.7 &      104.5 &      204.2&          24 &  0.10 \\ 
234 &  EPC &     189.4 &   10.8 &  10.0 &  21.50 &      0.745 &      0.586 &        4.6 &      268.8 &      189.4&         150 &  0.02 \\ 
235 &  LCY &     188.2 &  334.0 &  30.3 &  17.12 &      0.696 &      0.832 &       17.3 &      232.8 &      188.2&          35 &  0.02 \\ 
237 &  SSA &     192.8 &   39.2 &  10.9 &  38.13 &      0.956 &      0.084 &       16.4 &      150.7 &       12.8&          14 &  0.19 \\ 
241 &  OUI &     204.2 &  256.8 &  75.8 &  31.03 &      0.685 &      0.991 &       51.9 &      183.0 &      204.2&         152 &  0.10 \\ 
243 &  ZCN &     216.3 &  102.2 &  15.7 &  64.05 &      0.889 &      0.491 &      164.0 &       94.4 &       36.3&         547 &  0.20 \\ 
244 &  PAR &     222.5 &   97.4 &  53.9 &  54.16 &      0.944 &      0.520 &      107.7 &      267.9 &      222.5&           8 &  0.14 \\ 
245 &  NHD &     225.9 &  124.3 &  -2.7 &  65.84 &      0.840 &      0.928 &      139.8 &       31.0 &       46.2&          41 &  0.13 \\ 
250 &  NOO &     243.4 &   88.1 &  15.4 &  42.54 &      0.986 &      0.105 &       25.3 &      143.0 &       63.4&         233 &  0.05 \\ 
256 &  ORN &     251.7 &   75.8 &  25.5 &  25.25 &      0.783 &      0.450 &        2.6 &      284.2 &      251.7&         439 &  0.05 \\ 
257 &  ORS &     254.5 &   74.8 &  15.5 &  21.95 &      0.729 &      0.562 &        5.1 &       91.9 &       74.5&         208 &  0.06 \\ 
258 &  DAR &     265.0 &   75.6 &  29.6 &  18.57 &      0.709 &      0.689 &        3.8 &      253.3 &      265.0&          41 &  0.08 \\ 
260 &  GTI &     260.1 &   41.4 &  32.7 &  12.83 &      0.647 &      0.890 &        5.6 &      220.3 &      260.1&           6 &  0.12 \\ 
267 &  JNO &     291.4 &   83.7 &   1.3 &  14.24 &      0.634 &      0.862 &        8.1 &       46.2 &      111.4&          14 &  0.12 \\ 
268 &  BCD &     318.2 &  124.4 &   9.1 &  16.68 &      0.655 &      0.774 &        5.4 &       62.1 &      138.2&          54 &  0.06 \\ 
269 &  OCS &     318.7 &  348.2 &  69.8 &  12.80 &      0.577 &      0.984 &       17.3 &      178.7 &      318.7&           5 &  0.14 \\ 
273 &  PBO &      45.8 &  221.9 &  67.0 &  15.23 &      0.523 &      1.006 &       23.0 &      186.9 &       45.8&          13 &  0.14 \\ 
278 &  MSR &     107.3 &  236.9 &  24.3 &  11.46 &      0.582 &      1.005 &       11.9 &      194.3 &      107.3&          21 &  0.11 \\ 
279 &  ZED &     107.5 &  257.2 &  62.2 &  20.77 &      0.617 &      1.014 &       33.1 &      182.4 &      107.5&         121 &  0.11 \\ 
281 &  OCT &     192.6 &  166.7 &  78.6 &  45.01 &      0.935 &      0.991 &       77.5 &      168.8 &      192.6&          23 &  0.06 \\ 
282 &  DCY &     189.1 &  301.3 &  44.5 &  15.50 &      0.639 &      0.984 &       21.6 &      197.4 &      189.1&          36 &  0.13 \\ 
286 &  FTA &     232.1 &   59.0 &  14.7 &  24.52 &      0.767 &      0.460 &        5.1 &      103.8 &       52.1&         469 &  0.12 \\ 
288 &  DSA &     252.7 &   61.8 &   9.6 &  18.78 &      0.666 &      0.684 &        6.5 &       75.6 &       72.7&          28 &  0.13 \\ 
289 &  DNA &     244.6 &   43.0 &  24.4 &  15.26 &      0.665 &      0.792 &        2.9 &      239.8 &      244.6&          11 &  0.11 \\ 
323 &  XCB &     293.3 &  250.2 &  28.7 &  43.11 &      0.697 &      0.743 &       74.5 &      112.4 &      293.3&           5 &  0.13 \\ 
327 &  BEQ &      95.5 &  309.3 &   0.3 &  32.31 &      0.858 &      0.161 &       45.3 &      330.4 &       95.5&          19 &  0.19 \\ 
333 &  OCU &     202.3 &  145.4 &  64.1 &  54.34 &      0.858 &      0.977 &      100.3 &      163.1 &      202.3&         220 &  0.02 \\ 
348 &  ARC &      33.3 &  317.1 &  46.6 &  40.94 &      0.872 &      0.857 &       68.8 &      133.9 &       33.3&          87 &  0.08 \\ 
362 &  JMC &      73.9 &   16.2 &  54.1 &  42.92 &      0.925 &      0.611 &       68.4 &       99.0 &       73.9&           7 &  0.07 \\ 
365 &  BCM &     103.8 &   49.6 &  58.7 &  43.45 &      0.936 &      0.554 &       69.6 &       91.9 &      103.8&           5 &  0.14 \\ 
386 &  OBC &     202.9 &   52.5 &  56.6 &  49.31 &      0.921 &      0.494 &       84.5 &      272.9 &      202.9&          17 &  0.24 \\ 
388 &  CTA &     217.7 &   60.5 &  26.0 &  39.76 &      0.974 &      0.097 &       16.3 &      326.1 &      217.7&         141 &  0.08 \\ 
390 &  THA &     235.0 &   88.3 &  35.8 &  31.48 &      0.859 &      0.133 &       27.7 &      329.8 &      235.0&           5 &  0.05 \\ 
391 &  NDD &     234.9 &  258.6 &  64.2 &  26.28 &      0.655 &      0.988 &       42.3 &      176.7 &      234.9&          13 &  0.16 \\ 
392 &  NID &     232.9 &  200.2 &  65.1 &  42.28 &      0.742 &      0.985 &       74.9 &      173.4 &      232.9&          32 &  0.15 \\ 
401 &  BSX &     286.0 &  159.1 &   1.2 &  54.13 &      0.931 &      0.099 &      147.4 &      149.8 &      106.0&           7 &  0.10 \\ 
403 &  CVN &     286.4 &  199.5 &  44.0 &  52.87 &      0.885 &      0.860 &       94.3 &      223.1 &      286.4&          12 &  0.13 \\ 
404 &  GUM &     283.7 &  240.6 &  60.9 &  33.21 &      0.718 &      0.982 &       55.2 &      182.5 &      283.7&          18 &  0.27 \\ 
427 &  FED &     315.0 &  237.1 &  60.7 &  34.59 &      0.874 &      0.971 &       55.3 &      194.6 &      315.0&           7 &  0.13 \\ 
428 &  DSV &     267.0 &  203.9 &   4.9 &  65.24 &      0.916 &      0.574 &      150.7 &       96.8 &      267.0&          78 &  0.07 \\ 
430 &  POR &     173.5 &   72.5 &   6.6 &  65.88 &      0.857 &      0.860 &      151.1 &       47.2 &      349.8&         184 &  0.13 \\ 
431 &  JIP &      93.9 &  331.1 &  29.0 &  57.70 &      0.883 &      0.886 &      111.9 &      224.7 &       93.9&          32 &  0.11 \\ 
432 &  NBO &     296.4 &  200.6 &  12.3 &  65.89 &      0.811 &      0.837 &      144.4 &      232.4 &      296.4&          11 &  0.15 \\ 
434 &  BAR &     136.4 &   29.4 &  26.2 &  66.13 &      0.765 &      0.912 &      155.4 &      222.4 &      136.4&          43 &  0.11 \\ 
435 &  MPR &     140.1 &   70.3 &  51.2 &  56.29 &      0.827 &      0.607 &      119.4 &       91.1 &      140.1&          28 &  0.14 \\ 
436 &  GCP &     227.8 &   52.6 &  82.4 &  35.34 &      0.856 &      0.836 &       55.3 &      227.3 &      227.8&          18 &  0.14 \\ 
437 &  NLY &     228.7 &  135.0 &  41.2 &  62.55 &      0.708 &      0.818 &      135.6 &      234.8 &      228.7&          13 &  0.17 \\ 
438 &  MLE &     231.1 &  138.8 &  32.7 &  66.77 &      0.789 &      0.911 &      151.2 &      218.4 &      231.1&          40 &  0.10 \\ 
439 &  ASX &     235.0 &  155.4 &   1.2 &  69.64 &      0.866 &      0.914 &      164.8 &      325.7 &       55.0&           7 &  0.18 \\ 
440 &  NLM &     230.7 &  154.6 &  34.6 &  65.98 &      0.797 &      0.980 &      140.0 &      168.9 &      230.7&         105 &  0.11 \\ 
441 &  NLD &     232.3 &  185.6 &  73.2 &  40.54 &      0.651 &      0.969 &       70.8 &      199.1 &      232.3&          50 &  0.10 \\ 
442 &  RLE &     263.9 &  154.8 &   3.0 &  66.26 &      0.793 &      0.691 &      165.6 &       67.7 &       83.9&           8 &  0.09 \\ 
443 &  DCL &     253.5 &  146.4 &  26.9 &  64.43 &      0.875 &      0.576 &      153.8 &      264.4 &      253.5&          84 &  0.14 \\ 
444 &  ZCS &     112.5 &    9.8 &  50.3 &  56.08 &      0.891 &      0.999 &      106.2 &      164.3 &      112.5&         258 &  0.06 \\ 
450 &  AED &      20.5 &  308.4 &  11.4 &  59.93 &      0.935 &      0.710 &      120.7 &      113.0 &       20.5&          25 &  0.08 \\ 
451 &  CAM &      25.3 &  222.9 &  74.8 &  16.06 &      0.574 &      1.001 &       25.0 &      180.6 &       25.3&          28 &  0.17 \\ 
452 &  TVI &      32.3 &  203.1 &   4.1 &  16.17 &      0.649 &      0.793 &        6.0 &      242.3 &       32.3&          54 &  0.11 \\ 
454 &  MPV &      30.7 &  206.8 &   2.4 &  20.18 &      0.721 &      0.694 &        7.8 &      254.8 &       30.7&          62 &  0.12 \\ 
455 &  MAC &      41.6 &  209.1 &   6.4 &  15.53 &      0.654 &      0.848 &        7.8 &      233.4 &       41.6&          24 &  0.08 \\ 
456 &  MPS &      68.1 &  251.3 & -14.2 &  24.63 &      0.790 &      0.517 &        6.8 &      275.9 &       68.1&          12 &  0.08 \\ 
458 &  JEC &      82.6 &  314.8 &  33.2 &  51.85 &      0.889 &      0.909 &       95.3 &      219.1 &       82.6&          51 &  0.11 \\ 
459 &  JEO &      91.7 &  250.8 &  -1.2 &  14.88 &      0.631 &      0.877 &        7.7 &      229.8 &       91.7&          25 &  0.09 \\ 
460 &  LOP &      85.1 &  255.3 &  -2.9 &  18.86 &      0.708 &      0.743 &       11.0 &      249.0 &       85.1&          67 &  0.03 \\ 
463 &  JRH &     119.6 &  262.2 &  38.0 &  14.76 &      0.606 &      0.978 &       20.0 &      206.7 &      119.6&          37 &  0.06 \\ 
464 &  KLY &     125.8 &  271.4 &  37.5 &  18.00 &      0.675 &      0.966 &       24.3 &      208.8 &      125.8&          78 &  0.04 \\ 
465 &  AXC &     134.5 &   10.3 &  47.6 &  55.10 &      0.873 &      0.886 &      105.4 &      223.3 &      134.5&         144 &  0.04 \\ 
466 &  AOC &     142.4 &   34.2 &   2.6 &  65.63 &      0.912 &      0.707 &      158.4 &       69.3 &      322.4&          51 &  0.11 \\ 
468 &  AAH &     133.7 &  272.1 &  23.8 &  13.55 &      0.649 &      0.963 &       15.1 &      210.1 &      133.7&          33 &  0.11 \\ 
469 &  AUS &     144.0 &  287.6 &   0.1 &  10.83 &      0.603 &      0.947 &        6.3 &      213.2 &      144.0&          24 &  0.06 \\ 
470 &  AMD &     145.5 &  268.4 &  59.9 &  20.61 &      0.621 &      1.008 &       32.8 &      184.9 &      145.5&         657 &  0.05 \\ 
472 &  ATA &     145.8 &  313.2 &   0.3 &  18.19 &      0.694 &      0.749 &        9.5 &      248.1 &      145.8&         131 &  0.07 \\ 
473 &  LAQ &     155.4 &  351.3 &   1.0 &  31.06 &      0.876 &      0.252 &        6.5 &      308.9 &      155.4&         162 &  0.10 \\ 
474 &  ABA &     147.1 &  301.1 &   7.3 &  15.70 &      0.682 &      0.861 &       11.7 &      230.5 &      147.1&          93 &  0.04 \\ 
476 &  ICE &     170.3 &    5.5 &  -2.3 &  26.27 &      0.786 &      0.360 &        2.3 &      117.1 &      350.3&           5 &  0.10 \\ 
477 &  SRP &     169.3 &  341.1 &   3.6 &  18.28 &      0.709 &      0.709 &        6.0 &      253.4 &      169.3&         196 &  0.08 \\ 
479 &  SOO &     177.4 &   67.9 &  11.8 &  67.08 &      0.879 &      0.789 &      160.3 &       58.0 &      344.4&         102 &  0.23 \\ 
480 &  TCA &     202.2 &  132.4 &  29.6 &  66.62 &      0.795 &      0.809 &      158.6 &      124.8 &      202.2&         127 &  0.10 \\ 
481 &  OML &     215.6 &  145.0 &  31.2 &  66.90 &      0.813 &      0.889 &      150.7 &      140.2 &      215.6&          83 &  0.04 \\ 
485 &  NZT &     233.2 &   71.3 &  19.7 &  33.50 &      0.915 &      0.217 &        5.4 &      130.0 &       53.2&          59 &  0.08 \\ 
486 &  NZP &     231.1 &   57.4 &  29.7 &  27.87 &      0.844 &      0.380 &       10.9 &      292.7 &      231.1&          45 &  0.10 \\ 
488 &  NSU &     238.6 &  145.6 &  60.0 &  54.58 &      0.894 &      0.816 &       99.4 &      231.4 &      238.6&          22 &  0.10 \\ 
490 &  DGE &     258.5 &   75.9 &  -5.4 &  23.25 &      0.792 &      0.656 &       19.7 &       75.4 &       78.5&          11 &  0.14 \\ 
491 &  DCC &     249.7 &  129.7 &  11.6 &  62.35 &      0.917 &      0.390 &      165.3 &      107.1 &       69.7&          87 &  0.05 \\ 
495 &  DMT &     258.5 &   65.5 &   9.1 &  15.81 &      0.656 &      0.788 &        5.2 &       60.4 &       78.5&          49 &  0.07 \\ 
497 &  DAB &     262.0 &  210.8 &  22.1 &  59.54 &      0.964 &      0.684 &      114.2 &      112.2 &      262.0&           5 &  0.06 \\ 
500 &  JPV &     285.7 &  219.7 &   1.4 &  65.13 &      0.882 &      0.646 &      147.3 &      105.4 &      285.7&          51 &  0.08 \\ 
501 &  FPL &     317.7 &  149.4 &   7.9 &  28.53 &      0.829 &      0.363 &        5.2 &      114.1 &      137.7&          97 &  0.03 \\ 
502 &  DRV &     254.5 &  185.0 &  12.9 &  67.93 &      0.889 &      0.784 &      153.7 &      125.4 &      254.5&          63 &  0.04 \\ 
505 &  AIC &     133.8 &  344.6 & -14.8 &  38.64 &      0.956 &      0.104 &       22.6 &      147.1 &      313.8&         577 &  0.13 \\ 
506 &  FEV &     308.7 &  196.4 &  13.1 &  62.00 &      0.896 &      0.471 &      137.9 &      275.2 &      308.7&          65 &  0.12 \\ 
510 &  JRC &      84.7 &  321.4 &  44.4 &  49.13 &      0.886 &      1.007 &       87.6 &      190.9 &       84.7&          79 &  0.06 \\ 
511 &  FLY &     197.7 &   92.6 &  53.6 &  60.19 &      0.817 &      0.862 &      123.0 &      226.2 &      197.7&         175 &  0.10 \\ 
513 &  EPV &     253.0 &  192.6 &   8.8 &  65.23 &      0.935 &      0.538 &      150.4 &       93.6 &      253.0&          30 &  0.08 \\ 
515 &  OLE &     290.9 &  137.5 &   8.0 &  37.80 &      0.942 &      0.086 &       21.3 &      151.6 &      110.9&          26 &  0.07 \\ 
516 &  FMV &     322.2 &  218.6 &   4.2 &  65.49 &      0.852 &      0.718 &      143.8 &      245.2 &      322.2&          44 &  0.12 \\ 
517 &  ALO &      10.7 &  242.8 &   1.6 &  55.76 &      0.942 &      0.300 &      111.7 &      293.5 &       10.7&          13 &  0.11 \\ 
518 &  AHE &      32.4 &  271.1 &  20.6 &  51.80 &      0.878 &      0.780 &       93.9 &      239.8 &       32.4&          14 &  0.09 \\ 
519 &  BAQ &      45.0 &  322.7 &   0.1 &  67.84 &      0.886 &      0.921 &      155.4 &      144.8 &       45.0&          10 &  0.04 \\ 
522 &  SAP &     110.4 &  349.9 &  11.1 &  63.27 &      0.932 &      0.555 &      148.7 &      267.0 &      110.4&         260 &  0.04 \\ 
523 &  AGC &     155.4 &  344.1 &  75.9 &  43.47 &      0.843 &      1.004 &       75.0 &      188.7 &      155.4&         198 &  0.05 \\ 
524 &  LUM &     214.4 &  156.1 &  49.5 &  59.90 &      0.863 &      0.918 &      116.2 &      147.5 &      214.4&          83 &  0.07 \\ 
525 &  ICY &     223.6 &  286.5 &  56.8 &  18.64 &      0.654 &      0.989 &       28.2 &      183.9 &      223.6&          31 &  0.10 \\ 
526 &  SLD &     218.2 &  159.1 &  69.3 &  49.20 &      0.801 &      0.984 &       88.6 &      191.2 &      218.2&          85 &  0.09 \\ 
528 &  JZD &     275.9 &  241.2 &  69.4 &  29.34 &      0.614 &      0.973 &       47.1 &      191.6 &      275.9&          32 &  0.13 \\ 
529 &  EHY &     261.9 &  136.8 &   0.6 &  61.25 &      0.950 &      0.359 &      140.7 &      108.1 &       81.9&          91 &  0.09 \\ 
531 &  GAQ &      37.0 &  296.8 &  15.1 &  60.54 &      0.867 &      0.990 &      117.7 &      194.4 &       37.0&          13 &  0.19 \\ 
532 &  MLD &      50.9 &  202.9 &  81.3 &  15.48 &      0.605 &      1.000 &       22.5 &      168.2 &       50.9&          12 &  0.08 \\ 
533 &  JXA &     110.4 &   33.9 &   8.9 &  68.53 &      0.916 &      0.835 &      171.9 &      309.8 &      290.4&          71 &  0.10 \\ 
534 &  FOA &     135.3 &   24.2 &  47.1 &  60.65 &      0.839 &      1.003 &      121.3 &      192.2 &      135.3&          79 &  0.04 \\ 
537 &  KAU &     195.3 &   78.6 &  33.0 &  64.87 &      0.953 &      0.539 &      157.3 &      267.3 &      195.3&          66 &  0.14 \\ 
538 &  FFA &     212.6 &   51.2 &  28.8 &  36.79 &      0.942 &      0.216 &       21.0 &      308.0 &      212.6&          50 &  0.06 \\ 
539 &  ACP &     270.7 &  317.1 &  63.2 &  16.81 &      0.620 &      0.981 &       24.4 &      187.1 &      270.7&          18 &  0.12 \\ 
540 &  TCR &     271.7 &  166.6 &   0.9 &  69.35 &      0.892 &      0.802 &      171.3 &       52.5 &       91.7&           6 &  0.18 \\ 
541 &  SSD &     248.2 &  285.3 &  63.5 &  21.61 &      0.652 &      0.983 &       32.2 &      182.4 &      248.2&          39 &  0.12 \\ 
542 &  DES &     255.5 &  154.3 &   1.8 &  68.37 &      0.863 &      0.910 &      162.3 &       35.7 &       75.5&          13 &  0.13 \\ 
544 &  JNH &     273.6 &  148.1 &  -5.1 &  61.59 &      0.985 &      0.338 &      137.5 &      107.9 &       93.6&           7 &  0.25 \\ 
545 &  XCA &     155.1 &    8.2 &  49.0 &  50.60 &      0.914 &      0.695 &       93.9 &      250.3 &      155.1&          15 &  0.09 \\ 
546 &  FTC &     139.3 &   23.5 &  66.8 &  51.32 &      0.880 &      1.007 &       93.5 &      173.0 &      139.3&         142 &  0.07 \\ 
548 &  FAQ &     112.4 &  319.5 &  -2.0 &  37.04 &      0.926 &      0.129 &       35.0 &      325.3 &      112.4&         174 &  0.01 \\ 
549 &  FAN &     111.6 &   17.9 &  44.8 &  59.63 &      0.865 &      0.907 &      118.8 &      140.0 &      111.6&         214 &  0.08 \\ 
550 &  KPC &     114.4 &   26.0 &  59.4 &  51.16 &      0.881 &      0.993 &       92.4 &      161.2 &      114.4&          53 &  0.10 \\ 
551 &  FSA &     139.3 &   24.9 &  38.0 &  62.72 &      0.859 &      0.899 &      132.6 &      221.1 &      139.3&         134 &  0.07 \\ 
552 &  PSO &     161.0 &   71.2 &  -0.6 &  65.63 &      0.854 &      1.002 &      140.5 &       19.3 &      341.0&         103 &  0.12 \\ 
553 &  DPE &     164.8 &   54.2 &  44.0 &  63.74 &      0.854 &      0.888 &      135.5 &      221.9 &      164.8&          36 &  0.09 \\ 
554 &  APE &     160.9 &   39.1 &  45.5 &  60.98 &      0.883 &      0.798 &      126.0 &      236.2 &      160.9&         104 &  0.19 \\ 
555 &  OCP &     187.9 &   66.2 &  72.4 &  50.85 &      0.869 &      0.916 &       90.9 &      215.1 &      187.9&          35 &  0.11 \\ 
556 &  PTA &     186.5 &   57.4 &  28.3 &  59.60 &      0.960 &      0.251 &      156.8 &      303.3 &      186.5&          90 &  0.08 \\ 
557 &  SFD &     216.1 &  300.7 &  64.8 &  24.67 &      0.918 &      0.977 &       36.2 &      194.3 &      216.1&          53 &  0.06 \\ 
560 &  SES &     246.3 &  140.6 &   2.0 &  67.49 &      0.952 &      0.767 &      156.2 &       57.5 &       66.3&          10 &  0.25 \\ 
561 &  SSX &     254.9 &  139.0 &   0.9 &  65.48 &      0.930 &      0.616 &      149.9 &       77.5 &       74.9&          68 &  0.10 \\ 
562 &  BCT &     259.8 &  181.3 &  28.8 &  64.89 &      0.787 &      0.976 &      132.4 &      182.7 &      259.8&          31 &  0.13 \\ 
563 &  DOU &     267.8 &  159.7 &  42.9 &  56.36 &      0.937 &      0.542 &      106.6 &      266.0 &      267.8&          60 &  0.04 \\ 
564 &  SUM &     275.9 &  175.3 &  37.4 &  58.03 &      0.877 &      0.675 &      114.3 &      250.3 &      275.9&           8 &  0.15 \\ 
565 &  FUM &     273.5 &  169.8 &  43.5 &  56.21 &      0.944 &      0.631 &      105.1 &      254.3 &      273.5&          21 &  0.11 \\ 
566 &  BCF &     271.9 &  180.0 &  23.6 &  66.00 &      0.852 &      0.868 &      141.6 &      223.6 &      271.9&          38 &  0.12 \\ 
568 &  FCV &     306.4 &  196.4 &  30.7 &  55.12 &      0.919 &      0.594 &      102.0 &      262.1 &      306.4&          25 &  0.08 \\ 
570 &  FBH &     313.0 &  245.0 &  23.5 &  55.04 &      0.867 &      0.914 &      100.9 &      148.5 &      313.0&          30 &  0.07 \\ 
571 &  TSB &     343.1 &  216.2 &  24.3 &  49.18 &      0.956 &      0.502 &       82.4 &      270.4 &      343.1&          44 &  0.03 \\ 
572 &  TOH &     335.0 &  233.5 &   9.3 &  64.03 &      0.923 &      0.862 &      129.2 &      225.4 &      335.0&           5 &  0.25 \\ 
573 &  TLM &     250.0 &  164.0 &  38.6 &  63.26 &      0.826 &      0.929 &      127.8 &      209.3 &      250.0&          62 &  0.11 \\ 
574 &  GMA &     250.8 &  163.8 &  54.4 &  54.68 &      0.871 &      0.848 &      100.9 &      226.2 &      250.8&          44 &  0.08 \\ 
575 &  SAU &     233.0 &  105.4 &  38.7 &  55.27 &      0.969 &      0.190 &      119.6 &      310.8 &      233.0&          10 &  0.05 \\ 
576 &  FOB &     281.2 &  202.3 &  20.5 &  63.83 &      0.705 &      0.981 &      131.6 &      176.9 &      281.2&          10 &  0.18 \\ 
577 &  FPI &     128.7 &    9.1 &  12.4 &  63.74 &      0.954 &      0.483 &      161.9 &      275.4 &      128.7&          34 &  0.06 \\ 
578 &  TUM &     251.8 &  137.5 &  51.1 &  53.33 &      0.917 &      0.472 &      100.5 &      274.6 &      251.8&          29 &  0.08 \\ 
579 &  TCV &     278.6 &  208.7 &  29.5 &  59.79 &      0.791 &      0.963 &      114.0 &      161.6 &      278.6&          25 &  0.09 \\ 
580 &  CHA &     147.4 &   18.4 &  43.9 &  57.77 &      0.890 &      0.767 &      114.5 &      241.3 &      147.4&         118 &  0.13 \\ 
581 &  NHE &      30.2 &  261.2 &  39.4 &  39.51 &      0.906 &      0.899 &       64.4 &      216.9 &       30.2&          90 &  0.14 \\ 
583 &  TTA &     163.5 &   55.2 &   2.8 &  63.28 &      0.867 &      0.692 &      146.8 &       71.1 &      343.5&          39 &  0.09 \\ 
584 &  GCE &     137.4 &   24.9 &  74.6 &  45.84 &      0.849 &      0.976 &       81.1 &      156.1 &      137.4&         160 &  0.08 \\ 
586 &  TLA &     130.5 &  333.7 &  43.4 &  45.17 &      0.916 &      0.783 &       77.6 &      240.4 &      130.5&          85 &  0.08 \\ 
587 &  FNC &     130.5 &  313.3 &  45.5 &  35.88 &      0.893 &      0.840 &       57.4 &      230.3 &      130.5&          33 &  0.13 \\ 
588 &  TTL &     213.3 &  106.7 &  50.8 &  60.46 &      0.854 &      0.720 &      124.7 &      247.0 &      213.3&          37 &  0.12 \\ 
590 &  VCT &     277.3 &  180.6 &  45.1 &  53.72 &      0.900 &      0.771 &       98.2 &      238.3 &      277.3&          15 &  0.27 \\ 
591 &  ZBO &     319.4 &  215.1 &  17.4 &  61.30 &      0.904 &      0.741 &      121.6 &      244.0 &      319.4&          21 &  0.15 \\ 
592 &  PON &     135.5 &   16.7 &  28.2 &  64.12 &      0.895 &      0.771 &      142.1 &      240.4 &      135.5&         102 &  0.05 \\ 
593 &  TOL &     220.8 &  116.7 &  42.8 &  63.29 &      0.904 &      0.691 &      138.2 &      248.9 &      220.8&          48 &  0.06 \\ 
594 &  RSE &     290.7 &  230.0 &  21.7 &  57.36 &      0.951 &      0.816 &      106.1 &      130.4 &      290.7&          10 &  0.18 \\ 
595 &  TTT &     177.0 &   58.6 &  17.0 &  63.51 &      0.917 &      0.445 &      172.9 &      100.0 &      346.9&          77 &  0.09 \\ 
596 &  MUS &     276.7 &  190.7 &  59.2 &  44.47 &      0.874 &      0.847 &       76.2 &      226.8 &      276.7&          23 &  0.18 \\ 
599 &  POS &      16.8 &  283.6 &   7.2 &  63.88 &      0.948 &      0.998 &      128.4 &      186.3 &       16.8&           5 &  0.11 \\ 
600 &  FAU &     202.4 &   89.8 &  42.7 &  62.90 &      0.885 &      0.621 &      140.3 &      257.8 &      202.4&          75 &  0.13 \\ 
603 &  FCR &     225.7 &  116.5 &  32.4 &  64.51 &      0.895 &      0.547 &      155.1 &      269.8 &      225.7&          53 &  0.10 \\ 
604 &  ACZ &     278.1 &  118.2 &  16.9 &  33.28 &      0.897 &      0.195 &        4.8 &      131.7 &       98.1&          39 &  0.04 \\ 
605 &  FHR &       8.8 &  260.0 &  45.8 &  39.33 &      0.887 &      0.962 &       63.8 &      203.3 &        8.8&          11 &  0.10 \\ 
606 &  JAU &     276.9 &  170.1 &  63.6 &  40.10 &      0.844 &      0.751 &       64.7 &      241.8 &      276.9&          16 &  0.20 \\ 
607 &  TBO &     296.3 &  215.0 &  25.2 &  60.10 &      0.800 &      0.976 &      117.2 &      190.4 &      296.3&          37 &  0.08 \\ 
608 &  FAR &     180.8 &   60.7 &  29.7 &  63.47 &      0.909 &      0.487 &      159.4 &      272.7 &      180.8&          34 &  0.16 \\ 
609 &  BOT &     277.5 &  185.7 &  34.2 &  60.46 &      0.866 &      0.863 &      118.6 &      221.7 &      277.5&          42 &  0.23 \\ 
610 &  SGM &     251.1 &   94.0 &  14.8 &  41.47 &      0.985 &      0.132 &       23.4 &      138.6 &       71.1&         310 &  0.17 \\ 
611 &  VCF &     261.8 &  187.0 &  44.0 &  55.42 &      0.805 &      0.931 &      103.6 &      208.1 &      261.8&          17 &  0.15 \\ 
612 &  NCA &     250.9 &  192.1 &  49.3 &  53.15 &      0.789 &      0.981 &       96.7 &      171.2 &      250.9&          28 &  0.08 \\ 
613 &  TLY &     197.7 &  122.6 &  43.5 &  64.63 &      0.788 &      0.960 &      138.7 &      156.0 &      197.7&          67 &  0.10 \\ 
615 &  TOR &     285.6 &  186.8 &  24.6 &  62.78 &      0.871 &      0.791 &      129.6 &      234.4 &      285.6&          47 &  0.13 \\ 
616 &  TOB &     291.1 &  183.2 &  26.4 &  57.49 &      0.919 &      0.502 &      122.4 &      273.2 &      291.1&           5 &  0.18 \\ 
617 &  IUM &     223.9 &  132.3 &  50.6 &  60.67 &      0.787 &      0.888 &      122.5 &      220.0 &      223.9&          35 &  0.15 \\ 
619 &  SLM &     253.3 &  136.3 &  33.1 &  59.75 &      0.949 &      0.369 &      134.2 &      286.4 &      253.3&          26 &  0.10 \\ 
620 &  SBO &     295.3 &  204.9 &  31.5 &  56.65 &      0.815 &      0.883 &      109.4 &      219.7 &      295.3&           7 &  0.25 \\ 
621 &  SUA &     298.1 &  181.6 &  58.0 &  37.25 &      0.886 &      0.727 &       57.3 &      244.3 &      298.1&          37 &  0.16 \\ 
622 &  PUA &     231.7 &  129.0 &  55.0 &  57.23 &      0.895 &      0.774 &      110.7 &      237.1 &      231.7&          16 &  0.27 \\ 
  \hline
\end{longtable}